\pdfoutput=1
\documentclass[british]{scrartcl}
\usepackage[T1]{fontenc}
\usepackage[latin9]{inputenc}
\usepackage{geometry}
\usepackage{babel}
\usepackage{amsmath}
\usepackage{graphicx}
\usepackage{esint}
\usepackage[numbers]{natbib}
\usepackage[unicode=true,pdfusetitle,
 bookmarks=true,bookmarksnumbered=false,bookmarksopen=false,
 breaklinks=false,pdfborder={0 0 1},backref=false,colorlinks=false]
 {hyperref}

\makeatletter
\newcommand{\lyxaddress}[1]{
\par {\raggedright #1
\vspace{1.4em}
\noindent\par}
}

\usepackage{babel}
\usepackage{breakurl}

\usepackage{babel}
\AtBeginDocument{

}
\usepackage{color}
\newcommand{\red}{\color{black}}

\newcommand{\magenta}{\color{black}}

\makeatother

\begin{document}

\title{Finite size corrections in the random energy model and the replica
approach}

\author{Bernard Derrida\textsuperscript{1} and Peter Mottishaw\textsuperscript{2}}

\maketitle

\lyxaddress{\textsuperscript{1}Laboratoire de Physique Statistique, Ecole Normale
Sup\'{e}rieure, Universit\'{e} Pierre et Marie Curie, Universit\'{e} Paris Diderot,
CNRS, 24 rue Lhomond, 75231 Paris Cedex 05 - France}

\lyxaddress{\textsuperscript{2}SUPA, School of Physics and Astronomy, University
of Edinburgh, Mayfield Road, Edinburgh EH9 3JZ, United Kingdom}
\begin{abstract}
We present a  systematict and exact  way of computing finite size corrections for
the random energy model, in its low temperature phase. We obtain explicit
(though complicated) expressions for the finite size corrections of
the overlap functions. In its low temperature phase, the random energy
model is known to exhibit Parisi's broken symmetry of replicas. The
finite size corrections {\red given by our exact calculation can be
reproduced using replicas if we make specific assumptions about the fluctuations (with negative variances!) of the
number and sizes of the blocks when replica symmetry is broken.
As an alternative we  show that the exact expression for the non-integer moments
 of the partition function can be written in terms of coupled contour integrals over what can be thought of as 
  "complex replica numbers". Parisi's one step replica symmetry breaking arises naturally from the  saddle point of these integrals without
 making any ansatz  or using the replica method. 
 The fluctuations of the "complex replica numbers"   near the saddle point in the imaginary direction correspond to the negative 
 variances we observed in the replica calculation.  Finally} our approach allows one to see why some apparently diverging series or integrals
are harmless.
\end{abstract}

\section{Introduction}
Often the calculation of the extensive part of the free energy of
mean field models can be reduced to finding the saddle point of some
action which depends on an integer number (usually finite) of parameters
(e.g. the energy or the magnetization). Then fluctuations can be calculated
by replacing the action by its quadratic approximation near the saddle
point. Expanding around the saddle point also enables the finite size
corrections to be obtained. These are well known procedures which
work well as long as the number of variables, on which the saddle
point is calculated, is an integer. When one tries to apply the same
ideas to the theory of disordered systems using the replica approach,
the number of replicas is usually not an integer any more and the
first difficulty one has to face is to give a meaning to a quadratic
form with a non-integer number of variables. The difficulty is even
worse when the symmetry between this non-integer number of variables
is broken as in Parisi's theory of mean field spin glasses.

In 1979-1980 Parisi \citep{parisi_sequence_1980,parisi_order_1980,mezard_spin_1987}
proposed a replica based solution of the Sherrington-Kirkpatrick \citep{sherrington_solvable_1975,kirkpatrick_infinite-ranged_1978}
mean field model of spin glasses. In Parisi's theory, the extensive
part of the free energy could be determined by finding a saddle point
in an unusual domain: it was a saddle point in the space of $n\times n$
matrices where the size $n$ of the matrix was a continuous variable
(in fact in the replica calculation one had to take the limit $n\to0$
at the end of the calculation). Parisi was able to give a meaning
to such a saddle point when $n$ is not an integer.

Even before Parisi's work, for non-integer $n$, the Gaussian form
around the saddle point was already understood in the replica symmetric
phase by de Almeida and Thouless \citep{de_almeida_stability_1978}.
However, when replica symmetry is broken, determining the quadratic
form around the saddle point is far from obvious \citep{de_dominicis_eigenvalues_1983}.
This is why the form of the leading finite size corrections has been
debated for a long time and has made it difficult to connect the theory
with the results of numerical simulations \citep{parisi_critical_1993,boettcher_simulations_2010,parisi_several_1993,young_low-temperature_1982,young_direct_1983,palassini_ground-state_2008,billoire_numerical_2006,aspelmeier_free-energy_2008,aspelmeier_finite-size_2008}.
Understanding the fluctuations near the saddle point is also a necessary
step to build a field theory in finite dimension \citep{de_dominicis_spin_1984,de_dominicis_gaussian_1985}.

In the present paper, we present a full analysis of the fluctuations
near a saddle point with a broken replica symmetry for the random
energy model, a spin glass model much simpler than the Sherrington
Kirkpatrick model. Random energy models (REM) can be solved exactly
\citep{derrida_random-energy_1980,derrida_random-energy_1981,olivieri_existence_1984,kistler_derridas_2014}
without recourse to the replica method. But they can also be solved
using replicas and they are among the simplest models for which Parisi's
replica symmetry breaking \citep{parisi_sequence_1980,parisi_order_1980}
scheme holds \citep{derrida_random-energy_1981,gross_simplest_1984}.
However, computing finite size corrections using replicas has proved
challenging even for simple models such as the REM  \citep{campellone_non-perturbative_1995,campellone_replica_2009,dotsenko_replica_2011}.
We present here a systematic and direct way of computing the finite
size corrections of random energy models in the broken replica symmetry
phase. {\red Although  our calculations are done without recourse to replicas, our results} can be interpreted as due to fluctuations
of the parameters necessary to describe the broken replica symmetry.

Here we work with a Poisson version of the REM, which is exponentially
close, for large system sizes, to the original REM (see Appendix A).
How this Poisson REM is defined and how the overlaps or the moments
of the partition function can be computed for this Poisson REM is
the purpose of section 2. In section 3, we develop a systematic way
of computing the finite size corrections of the overlaps and of the
non-integer moments of the partition function. In section 4, we discuss
how the results of section 3, for the finite size corrections of the
overlaps, can be {magenta reproduced using}  Parisi's broken
replica symmetry  {magenta scheme}. We show that to obtain the correct finite size corrections,
one has to supplement Parisi's ansatz by fluctuations of the replica
numbers, with negative variances. In section 5, 
{\red we write exact expressions of}
the non-integer moments of the partition function as contour integrals
over "complex replica numbers" (\ref{XXX2}). Then Parisi's ansatz appears
 {\magenta naturally} as the saddle point in these replica numbers, and the fluctuations
calculated at this saddle point are consistent with the fluctuations
predicted in the previous sections.

\section{A Poisson process version of the random energy model}

\label{poisson}

In this section we first recall a few known results on the random
energy model. We then define a Poisson process version of the REM,
for which we show how to compute the overlaps and the non-integer
moments of the partition function. It is known that in the low temperature
phase of the REM, one can represent the energies by a Poisson process
{\red\citep{ruelle_mathematical_1987,bovier_fluctuations_2002,kratz_representation_2004}}. While
in the large $N$ limit it is sufficient to take a Poisson process
with an exponential density, here, because we are interested by finite
size effects, we need to include corrections to this exponential density.

\subsection{The random energy model (REM) }

In the random energy model, one considers a system with $2^{N}$ possible
configurations ${\cal C}$, the energies $E({\cal C})$ of which are
i.i.d. random variables distributed according to a probability distribution
\begin{equation}
P(E({\cal C}))=\frac{1}{\sqrt{\pi NJ^{2}}}\exp\left\{ -\frac{E({\cal C})^{2}}{NJ^{2}}\right\} \ .\label{dist_rem}
\end{equation}
A sample is characterized by the choice of these $2^{N}$ random energies
$E({\cal C})$ and as usual in the theory of disordered systems the
first quantity of interest is, for a typical sample, the free energy
$F=\log Z(\beta)$ where 
\[
Z(\beta)=\sum_{{\cal C}}e^{-\beta E({\cal C})}\ \ \ \ \ \ \ \text{with}\ \ \ \beta=1/T\ .
\]
One of the remarkable features of the REM is that, in the large $N$
limit, it undergoes a freezing transition at a critical temperature
\citep{derrida_random-energy_1980,derrida_random-energy_1981}

\begin{equation}
T_{c}=J/(2\sqrt{\ln2})\label{eq:trsn_temp}
\end{equation}
and that, below this temperature, the partition function is dominated
by the energies of the configurations close to the ground state \citep{derrida_random-energy_1980,derrida_random-energy_1981}
\begin{equation}
E_{{\rm ground\ state}}\simeq-NJ\sqrt{\ln2}\ .\label{EGS}
\end{equation}

The REM is the simplest spin glass model which exhibits broken replica
symmetry \citep{derrida_random-energy_1981,gross_simplest_1984}:
the overlap $q({\cal C},{\cal C}')$ between two configurations can
take only two possible values $0$ or $1$ 
\[
q({\cal C},{\cal C}')=\delta_{{\cal C},{\cal C}'}
\]
and the Parisi's function $q(x)$ is a step function \citep{gross_simplest_1984,mezard_spin_1987}
\begin{equation}
q(x)=\Theta(x-1+\langle P_{2}\rangle)\label{q-parisi}
\end{equation}
where $\Theta(x)$ is the Heaviside function, $P_{2}$ is the probability
of finding, at equilibrium, two copies of the same sample in the same
configuration 
\begin{equation}
P_{2}=\sum_{{\cal C}}\left(\frac{e^{-\beta E({\cal C})}}{\sum_{{\cal C}}e^{-\beta E({\cal C})}}\right)^{2}=\frac{Z(2\beta)}{Z(\beta)^{2}}\label{P_2_def}
\end{equation}
and $\langle.\rangle$ in (\ref{q-parisi}) denotes an average over
the samples, i.e. over the random energies $E({\cal C})$.

In the large $N$ limit, $P_{2}$ vanishes in the high temperature
phase, while at low temperature (in the frozen phase) it takes non
zero values with sample to sample fluctuations, because it is dominated
by the ground state and the lowest excited states. In the $N\to\infty$
limit, direct calculations as well as replica calculations have shown
\citep{gross_simplest_1984,derrida_sample_1985} that, below $T_{c}$,
\begin{equation}
\langle P_{2}\rangle=1-\mu\ \ \ \ \ \text{with}\ \ \ \mu=\frac{T}{T_{c}}=\frac{2\sqrt{\log2}}{\beta J}\label{P2-infinite}
\end{equation}
One of the goals of the present paper is to present a method to calculate
the finite size corrections to this result in order to understand
the effect of fluctuations in the space of replicas.

The quantity $P_{2}$ (which is nothing but the thermal average of
the overlap $q({\cal C},{\cal C}')$) can be generalized to the probabilities
$P_{k}$ of finding $k$ copies of the same sample in the same configuration
\begin{equation}
P_{k}=\sum_{{\cal C}}\left(\frac{e^{-\beta E({\cal C})}}{\sum_{{\cal C}}e^{-\beta E({\cal C})}}\right)^{k}=\frac{Z(k\beta)}{Z(\beta)^{k}}\label{P_k_def}
\end{equation}
and generalized further to the probabilities $P_{k_{1},\cdots,k_{p}}$
of finding $k_{1}$ copies in the same configuration, $k_{2}$ copies
in a different configuration, $\cdots$, $k_{p}$ in yet another configuration
\begin{equation}
P_{k_{1},\cdots,k_{p}}=\frac{\sum_{{\cal C}_{1},\cdots{\cal C}_{p}}e^{-\beta(k_{1}E({\cal C}_{1})+\cdots k_{p}E({\cal C}_{p}))}}{\left(\sum_{{\cal C}}e^{-\beta E({\cal C})}\right)^{k_{1}+\cdots k_{p}}}\label{P_kp_def}
\end{equation}
where in the numerator of (\ref{P_kp_def}), the sum is over all possible
sets of $p$ different configurations ${\cal C}_{1}\cdots{\cal C}_{p}$.
As for $P_{2}$, the large $N$ limits of the averages of these overlaps
are known \citep{mezard_nature_1984,mezard_replica_1984,mezard_spin_1987,derrida_random_1997}
\begin{equation}
\langle P_{k}\rangle=\frac{\Gamma(k-\mu)}{\Gamma(1-\mu)\ \Gamma(k)}\ \ \ ;\ \ \ \ \ \ \langle P_{k_{1},\cdots,k_{p}}\rangle=\mu^{p-1}\frac{\Gamma(p)}{\Gamma(k_{1}+\cdots k_{p})}\ \frac{\Gamma(k_{1}-\mu)}{\Gamma(1-\mu)}\cdots\ \frac{\Gamma(k_{p}-\mu)}{\Gamma(1-\mu)}\label{Pk-infinite}
\end{equation}
and we will discuss below how to calculate their finite size corrections.

\subsection{A Poisson process version of the REM \label{sub:Definition-of-Poisson-REM}}

To slightly simplify the discussion below, we consider in the present
paper a Poisson process version of the REM, \textit{the Poisson REM}.
{\red The same idea of replacing a random energy model by a Poisson process was already used \cite{ogure_exact_2004} for the DREM (a  version of the REM where energies can take only integer values)}.
In this Poisson REM, the values of the energies are the points generated
by a Poisson process on the real line with intensity 
\begin{equation}
\rho\left(E\right)=\frac{2^{N}}{\sqrt{\pi NJ^{2}}}\exp\left\{ -\frac{E^{2}}{NJ^{2}}\right\} \ .\label{eq:energy_dist}
\end{equation}
This means that each infinitesimal interval $(E,E+\Delta E)$ on the
real line is either empty, with probability $1-\rho(E)\Delta E$,
or occupied by a single configuration with probability $\rho(E)\Delta E$.
As we eventually take \ensuremath{\Delta E\rightarrow0} it is justified
to forget events where more than one level falls into the interval
$(E,E+\Delta E)$.

One way of thinking of this Poisson REM is to divide the energy axis
into intervals of size \ensuremath{\Delta E} and label each interval
with an integer \ensuremath{j\in(-\infty,+\infty)}. The energy associated
with interval \ensuremath{j} is given by \ensuremath{j\Delta E}.
A realisation of the disorder is given by a set of independent random
binary variables \ensuremath{\{y_{j}\}}, which determine if the interval
\ensuremath{j} contains an energy level 
\begin{equation}
y_{j}=\begin{cases}
1 & \textnormal{if interval}\: j\:\textnormal{contains an energy level,}\\
0 & \textnormal{ if not\ . }
\end{cases}\label{eq:first}
\end{equation}
These independent random variables \ensuremath{\{y_{j}\}} are chosen
according to 
\begin{equation}
y_{j}=\begin{cases}
1\,\,\textnormal{with probability}\: & \rho\left(j\Delta E\right)\Delta E\\
0\,\,\textnormal{with probability}\: & 1-\rho\left(j\Delta E\right)\Delta\ .\ 
\end{cases}\label{eq:yj_dstn}
\end{equation}

The partition function, for a particular realisation \ensuremath{\{y_{j}\}}
of disorder, is given by

\begin{equation}
Z\left[\left\{ y_{j}\right\} \right]=\sum_{j=-\infty}^{+\infty}y_{j}e^{-\beta j\Delta E}.\label{eq:second}
\end{equation}

and the probability, at equilibrium, of finding the system in a specific
energy interval is 
\begin{equation}
\Pr(\textrm{system in interval \ensuremath{i}})=\frac{y_{i}\;\mathrm{e}^{-\beta i\Delta E}}{\sum\limits _{j=-\infty}^{+\infty}y_{j}\;\mathrm{e}^{-\beta j\Delta E}}.\label{eq:state_prby}
\end{equation}

The Poisson REM has on average $2^{N}$ energy levels. In the large
$N$ limit, it has the same free energy as the REM (see appendix \ref{sec:Poisson_rem}),
with in particular the same transition temperature (\ref{eq:trsn_temp}).
One difference, though, is that the total number of configurations
fluctuates in the Poisson REM while it is fixed in the REM.

As for the REM, the low temperature phase of the Poisson REM is dominated
by the energy levels close to the ground state and the average overlaps
are given by (\ref{Pk-infinite}) in the large $N$ limit. In fact,
as shown in appendix A, the difference between the free energies of
REM and of the Poisson REM is exponentially small in the system size
$N$. So we expect all the $1/N$ corrections to be the same for both
models.

\subsection{Expressions of the overlaps in the Poisson REM}

In the Poisson REM, the probabilities $P_{k}$ defined in (\ref{P_k_def})
take the form (\ref{eq:state_prby})

\begin{equation}
P_{k}=\sum_{i=-\infty}^{+\infty}\left[\Pr(\textrm{system in interval \ensuremath{i}})\right]^{k}=\frac{1}{Z\left[\left\{ y_{j}\right\} \right]^{k}}\sum_{i=-\infty}^{+\infty}y_{i}\mathrm{e}^{-\beta ki\Delta E}.\label{eq:pk_sums}
\end{equation}

One difficulty when one tries to average (\ref{eq:pk_sums}) over
the \ensuremath{\{y_{i}\}} is the presence of \ensuremath{Z} in
the denominator. This difficulty can be overcome by using an integral
representation of the Gamma function 
\begin{align}
\frac{1}{{Z}^{k}} & =\frac{1}{\Gamma(k)}\int_{0}^{\infty}\mathrm{d}t\, t^{k-1}\:\mathrm{e}^{-{Z}t}=\frac{1}{\Gamma(k)}\int_{0}^{\infty}\mathrm{d}t\: t^{k-1}\prod_{{j}=-\infty}^{\infty}\left[\exp{\left(-ty_{{j}}\mathrm{e}^{-\beta{j}\Delta E}\right)}\right].\label{int-repr}
\end{align}

Then using (\ref{eq:yj_dstn}) to average (\ref{eq:pk_sums}) over
the \ensuremath{\{y_{i}\}} and taking the limit $\Delta E\to0$ gives
\begin{equation}
\langle P_{k}\rangle=\frac{1}{\Gamma(k)}\int_{0}^{\infty}\mathrm{d}t\: t^{k-1}F_{k}(t)\exp{(-F(t)}),\label{eq:pk_average_integral}
\end{equation}

where 
\begin{equation}
F(t)=\int_{-\infty}^{+\infty}\left(1-\exp{\left(-t\mathrm{e}^{-\beta E}\right)}\right)\rho\left(E\right)\mathrm{d}E\label{eq:f_integral}
\end{equation}
and 
\begin{equation}
F_{k}(t)=(-)^{k{+1}}\frac{d^{k}F(t)}{\mathrm{d}t^{k}}=\int_{-\infty}^{+\infty}{\exp\left(-k\beta E-t\mathrm{e}^{-\beta E}\right)}\rho\left(E\right)\mathrm{d}E\,.\label{eq:fk_integral}
\end{equation}

A similar calculation for the more general overlap (\ref{P_kp_def})
leads to 
\begin{equation}
\langle P_{k_{1},\cdots k_{p}}\rangle=\frac{1}{\Gamma(k_{1}+\cdots k_{p})}\int_{0}^{\infty}\mathrm{d}t\: t^{k_{1}+\cdots k_{p}-1}F_{k_{1}}(t)\cdots F_{k_{p}}(t)\exp{(-F(t)})\,.\label{eq:pkp_average_integral}
\end{equation}

Expressions (\ref{eq:pk_average_integral}-\ref{eq:pkp_average_integral})
(and (\ref{Znu}-\ref{eq:pkp_average_integral-nu}) below) are exact
for a Poisson REM with an arbitrary $\rho(E)$.

\subsection{The moments of $Z$ and the weighted overlaps}

As we will discuss below for the replica approach, it is also useful
to obtain exact expressions for the moments of the partition function
and for the weighted overlaps which will appear in the replica approach
of section 4. Expressions of the integer or non-integer moments \citep{kondor_parisis_1983,gardner_probability_1989}
of $Z$ are useful to calculate the fluctuations and the large deviations
of the free energy \citep{dorlas_large_2001}.To compute the non-integer
moments of the partition function $\langle Z^{m}\rangle$ for $0<m<1$
we use again an integral representation similar to (\ref{int-repr})
\begin{equation}
\langle Z^{m}\rangle=\frac{1}{\Gamma(-m)}\int_{0}^{\infty}\mathrm{d}t\: t^{-m-1}\Big(\left\langle e^{-tZ}\right\rangle -1\Big)\ .\label{Znu0}
\end{equation}
(the calculation below could be extended to $m\notin(0,1)$ by replacing
(\ref{Znu0}) by the appropriate representation of the Gamma function.)

By averaging over the \ensuremath{\{y_{i}\}} as above in (\ref{int-repr},\ref{eq:pk_average_integral})
one gets
{\red an  expression  very similar to expression (3.8) of Ogure and Kabashima \cite{ogure_exact_2004} for the DREM}
\begin{equation}
\langle Z^{m}\rangle=\frac{1}{\Gamma(-m)}\int_{0}^{\infty}\mathrm{d}t\: t^{-m-1}\Big(\exp{(-F(t)})-1\Big)\, . \label{Znu}
\end{equation}
One can also define generalized weighted overlaps (where events are
weighted by powers of the partition function)

\begin{equation}
\langle P_{k}\rangle_{m}=\frac{\left\langle P_{k}Z^{m}\right\rangle }{\langle Z^{m}\rangle}\ \ \ \ \ \ ;\ \ \ \ \ \ \langle P_{k_{1},\cdots k_{p}}\rangle_{m}=\frac{\left\langle P_{k_{1},\cdots k_{p}}Z^{m}\right\rangle }{\langle Z^{m}\rangle}\,.\label{general-overlap}
\end{equation}
Using for $k\ge1$ and $0<m<1$ the identity 
\[
Z^{m-k}=\frac{1}{\Gamma(k-m)}\int_{0}^{\infty}\mathrm{d}t\: t^{k-m-1}e^{-tZ}
\]
one gets 
\begin{equation}
\langle P_{k}Z^{m}\rangle=\frac{1}{\Gamma(k-m)}\int_{0}^{\infty}\mathrm{d}t\: t^{k-m-1}\; F_{k}(t)\;\exp{(-F(t)})\ ,\label{eq:pk_average_integral-nu}
\end{equation}
and 
\begin{equation}
\langle P_{k_{1},\cdots k_{p}}Z^{m}\rangle=\frac{1}{\Gamma{(k_{1}+\cdots k_{p}-m)}}\int_{0}^{\infty}\mathrm{d}t\: t^{k_{1}+\cdots k_{p}-m-1}\; F_{k_{1}}(t)\cdots F_{k_{p}}(t)\;\exp{(-F(t)})\ .\label{eq:pkp_average_integral-nu}
\end{equation}
Formulas (\ref{Znu}-\ref{eq:pkp_average_integral-nu}) are exact
for the Poisson REM. They summarize all the previous ones in particular
(\ref{eq:pk_average_integral},\ref{eq:pkp_average_integral}). For
example one recovers (\ref{eq:pkp_average_integral}) by taking the
$m\to0$ limit. They will be our starting point to calculate $1/N$
corrections.

\section{Finite size corrections to the overlap functions\label{sec:Finite-size-corrections}}

\subsection{A direct calculation of finite size corrections}

In the low temperature phase, the partition function of the REM or
of the Poisson REM is dominated by the energies close to the ground
state energy. It is therefore legitimate to replace the density $\rho(E)$
by an approximation valid in the neighbourhood of the ground state
energy. Let us write (\ref{eq:energy_dist}) as

\begin{equation}
\rho\left(E\right)=A\ \exp\left[\alpha(E-E_{0})-\epsilon(E-E_{0})^{2}\right]\label{eq:rho_in_epsilon_form0}
\end{equation}
where we define (\ref{EGS}) 
\begin{equation}
E_{0}=-NJ\sqrt{\log2}\ ,\label{E0-def}
\end{equation}
\begin{equation}
\alpha=\frac{2\left|E_{0}\right|}{NJ^{2}}=\frac{2\sqrt{\log2}}{J}\ ,\label{alpha-def}
\end{equation}
\begin{equation}
\epsilon=\frac{1}{NJ^{2}}\ ,\label{epsilon-def}
\end{equation}
and 
\begin{equation}
A=\frac{1}{\sqrt{\pi NJ^{2}}}\ .\label{eq:A_dfnn}
\end{equation}
In the REM, the distances between the energies of the ground state
and of the lowest excited states remain of order 1 (in the large $N$
limit) and $E_{{\rm ground\ state}}-E_{0}={\cal O}(\log N)$ (see
\citep{derrida_random-energy_1981,galves_fluctuations_1989,campellone_replica_2009})
so that (\ref{eq:rho_in_epsilon_form0}) is valid in the vicinity
of the ground state.

Note that with these definitions (\ref{alpha-def}), one has (\ref{P2-infinite})
\begin{equation}
\mu=\frac{\alpha}{\beta}\,.\label{mu-alpha-beta}
\end{equation}
Therefore for $\epsilon$ small (i.e. for large $N$), one can replace
(\ref{eq:rho_in_epsilon_form0}) by 
\begin{equation}
\rho\left(E\right)=A\ \exp\left[\alpha(E-E_{0})\right]\left(1-\epsilon(E-E_{0})^{2}+{\cal O}(\epsilon^{2})\right)\label{eq:rho_in_epsilon_form}
\end{equation}
and this can be written as 
\begin{equation}
\rho(E)=A\left.\left(1-\epsilon\frac{{d}^{2}}{{d}\gamma^{2}}\right)e^{\gamma(E-E_{0})}\right|_{\gamma=\alpha}+\mathcal{O}\left(\epsilon^{2}\right),\label{rho-new}
\end{equation}
so that (\ref{eq:f_integral},\ref{eq:fk_integral}) become 
\begin{equation}
F(t)=\frac{A}{\alpha}\ \Gamma\left(1-\frac{\alpha}{\beta}\right)\ t^{\frac{\alpha}{\beta}}e^{-\alpha E_{0}}-\left.\epsilon\frac{{d}^{2}}{{d}\gamma^{2}}\left(\frac{A}{\gamma}\ \Gamma\left(1-\frac{\gamma}{\beta}\right)\ t^{\frac{\gamma}{\beta}}e^{-\gamma E_{0}}\right)\right|_{\gamma=\alpha}+\mathcal{O}\left(\epsilon^{2}\right),\label{Ft}
\end{equation}
\begin{equation}
F_{k}(t)=\frac{A}{\beta}\ \Gamma\left(k-\frac{\alpha}{\beta}\right)\ t^{\frac{\alpha}{\beta}-k}e^{-\alpha E_{0}}-\left.\epsilon\frac{{d}^{2}}{{d}\gamma^{2}}\left(\frac{A}{\beta}\ \Gamma\left(k-\frac{\gamma}{\beta}\right)\ t^{\frac{\gamma}{\beta}-k}e^{-\gamma E_{0}}\right)\right|_{\gamma=\alpha}+\mathcal{O}\left(\epsilon^{2}\right).\label{Fkt}
\end{equation}
under the condition 
\begin{equation}
\alpha<\beta\ \ \ \ {\rm i.e.}\ \ \ \mu<1\ .\label{condition}
\end{equation}

Substituting the expansions (\ref{Ft},\ref{Fkt}) into the integral
form (\ref{eq:pk_average_integral}) of $\left\langle P_{k}\right\rangle $
gives, 
\begin{equation}
\left\langle P_{k}\right\rangle =\frac{\Gamma\left(k-\frac{\alpha}{\beta}\right)}{\Gamma\left(k\right)\Gamma\left(1-\frac{\alpha}{\beta}\right)}+\frac{\epsilon{A}}{\alpha\ \Gamma\left(k\right)\Gamma\left(1-\frac{\alpha}{\beta}\right)}\frac{{d^{2}}B(\gamma)}{{d}\gamma^{2}}\bigg|_{\gamma=\alpha}+\mathcal{O}\left(\epsilon^{2}\right)
\end{equation}

where 
\[
\mathrm{B\left(\gamma\right)}=\Gamma\left(\frac{\gamma}{\alpha}\right)\left[\frac{\alpha}{A\,\Gamma\left(1-\frac{\alpha}{\beta}\right)}\right]^{\frac{\gamma}{\alpha}}\left[\Gamma\left(1-\frac{\gamma}{\beta}\right)\Gamma\left(k-\frac{\alpha}{\beta}\right)-\Gamma\left(k-\frac{\gamma}{\beta}\right)\Gamma\left(1-\frac{\alpha}{\beta}\right)\right].
\]
This gives 
\begin{eqnarray}
\left\langle P_{k}\right\rangle  & = & \frac{\Gamma\left(k-\frac{\alpha}{\beta}\right)}{\Gamma\left(k\right)\Gamma\left(1-\frac{\alpha}{\beta}\right)}+\epsilon\left[\left(2\frac{\log(A\Gamma(1-\frac{\alpha}{\beta}))-\Gamma'(1)-\log\alpha}{\alpha}+2\frac{\Gamma'(1-\frac{\alpha}{\beta})}{{\beta}\Gamma(1-\frac{\alpha}{\beta})}\right)\frac{d}{d\alpha}\left(\frac{\Gamma\left(k-\frac{\alpha}{\beta}\right)}{\Gamma\left(k\right)\Gamma\left(1-\frac{\alpha}{\beta}\right)}\right)\right.\nonumber \\
 &  & \,\,\,\,\,\,\,\,\,\,\,\,\,\,\,\,\,\,\,\,\,\,\,\,\,\,\,\,\,\,\,\,\,\,\,\left.-\frac{d^{2}}{d\alpha^{2}}\left(\frac{\Gamma\left(k-\frac{\alpha}{\beta}\right)}{\Gamma\left(k\right)\Gamma\left(1-\frac{\alpha}{\beta}\right)}\right)\right]+\mathcal{O}\left(\epsilon^{2}\right)\,.\label{R1}
\end{eqnarray}

We recover the known \citep{mezard_nature_1984,mezard_replica_1984,derrida_sample_1985,mezard_spin_1987,derrida_random_1997}
zero-th order term (\ref{Pk-infinite}). By replacing $\alpha$ by
its expression (\ref{alpha-def})\ and by using the expression (\ref{mu-alpha-beta})
for $\mu$ one gets the $1/N$ correction 
\begin{equation}
\left\langle P_{k}\right\rangle =\frac{\Gamma\left(k-\mu\right)}{\Gamma\left(k\right)\Gamma\left(1-\mu\right)}+\frac{1}{N}\left[\Delta_{1}\frac{d}{d\mu}\left(\frac{\Gamma\left(k-\mu\right)}{\Gamma\left(k\right)\Gamma\left(1-\mu\right)}\right)+\Delta_{2}\frac{d^{2}}{d\mu^{2}}\left(\frac{\Gamma\left(k-\mu\right)}{\Gamma\left(k\right)\Gamma\left(1-\mu\right)}\right)\right]+o\left(\frac{1}{N}\right)\label{R2}
\end{equation}
where 
\[
\Delta_{1}=\frac{-\Gamma'(1)+\log(\Gamma(1-\mu))-\log(2\sqrt{N\pi\log2})}{2\log2}\mu+\frac{\mu^{2}}{2\log2}\frac{\Gamma'(1-\mu)}{\Gamma(1-\mu)}\,,
\]
\[
\Delta_{2}=-\frac{1}{4\log2}\mu^{2}\,.
\]

The idea introduced in this section to compute $1/N$ corrections
is straightforward enough to be extended to compute higher orders
or the finite size corrections of other quantites, like the moments
of the partition function (\ref{Znu}) or the weighted generalized
overlaps (\ref{general-overlap}, \ref{eq:pkp_average_integral-nu}).

\subsection{An alternative way of computing finite size corrections}

We discuss now an alternative way of computing the $1/N$ corrections
which is somewhat simpler. One can rewrite (\ref{eq:rho_in_epsilon_form})
as 
\begin{equation}
\rho(E)=\left\{ Ae^{(\alpha+\phi)(E-E_{0})}\right\} _{\phi}+{\cal O}(\epsilon^{2})\label{rho(E)2}
\end{equation}
where $\phi$ is a random variable (of order $\epsilon^{\frac{1}{2}}$)
which satisfies 
\begin{equation}
\left\{ \phi\right\} _{\phi}=0\ \ \ \ \ \ \ \ ;\ \ \ \ \ \ \ \left\{ \phi^{2}\right\} _{\phi}=-2\epsilon\label{phi-stat}
\end{equation}
and $\left\{ .\right\} _{\phi}$ denotes an average over the variable
$\phi$. Negative variances appear here and in several other places
in this paper. Here (\ref{phi-stat}) simply means that for an arbitrary
function $G(\phi)$ one has 
\begin{equation}
\left\{ G(\phi)\right\} _{\phi}=G(0)-\epsilon G''(0)+{\cal O}(\epsilon^{2})\label{Gphi}
\end{equation}
(alternatively one could think of $\phi$ as being a pure imaginary
random number).

Using (\ref{rho(E)2},\ref{phi-stat},\ref{Gphi}) in (\ref{eq:f_integral},\ref{eq:fk_integral})
one gets for $F(t)$ and the $F_{k_{i}}(t)$ 
\[
F(t)=-\frac{A}{\beta}\left\{ \Gamma\left(-\mu_{0}\right)\ t^{\mu_{0}}e^{-\beta\mu_{0}E_{0}}\right\} _{\phi_{0}}+{\cal O}(\epsilon^{2})
\]
\[
F_{k_{i}}(t)=\frac{A}{\beta}\left\{ \Gamma\left(k_{i}-\mu_{i}\right)\ t^{\mu_{i}-k_{i}}e^{-\beta\mu_{i}E_{0}}\right\} _{\phi_{i}}+{\cal O}(\epsilon^{2})
\]
where for $1\le i\le p$ 
\[
\mu_{i}=\frac{\alpha+\phi_{i}}{\beta}=\mu+\frac{\phi_{i}}{\beta}\ .
\]

As at order $\epsilon$ one has 
\[
e^{-F(t)}=e^{-F^{*}(t)}(1-F(t)+F^{*}(t))+{\cal O}(\epsilon^{2})
\]
where 
\[
F^{*}(t)=-\frac{A}{\beta}\ \Gamma\left(-\mu\right)\ t^{\mu}e^{-\beta\mu E_{0}}
\]

and this gives for the weighted overlaps (\ref{eq:pkp_average_integral})
using the fact that the difference $F-F^{*}$ is of order $\epsilon$
\begin{eqnarray}
\langle P_{k_{1},\cdots k_{p}}Z^{m}\rangle & = & \left(\frac{A}{\beta}\right)^{p}e^{-\beta mE_{0}}\left[\left\{ \frac{\Gamma(k_{1}-\mu_{1})\cdots\Gamma(k_{p}-\mu_{p})\ \Gamma(\frac{\mu_{1}+\cdots\mu_{p}-m}{\mu})}{\mu\ \Gamma{(k_{1}+\cdots k_{p}-m)}}\left(-\frac{A}{\beta}\Gamma\left(-\mu\right)\right)^{\frac{m-\mu_{1}-\cdots\mu_{p}}{\mu}}\right\} _{\phi_{1},\cdots,\phi_{p}}\right.\nonumber \\
 & + & \left(\frac{A}{\beta}\right)\frac{\Gamma(k_{1}-\mu)\cdots\Gamma(k_{p}-\mu)\ \Gamma(p+1-\frac{m}{\mu})}{\mu\ \Gamma{(k_{1}+\cdots k_{p}-m)}}\left(-\frac{A}{\beta}\Gamma\left(-\mu\right)\right)^{\frac{m}{\mu}-p-1}\left(-\Gamma(-\mu)\right)\nonumber \\
 & - & \left.\vphantom{\left\{ \left(-\frac{A}{\beta}\Gamma\left(-\mu\right)\right)^{\frac{m-\mu_{1}-\cdots\mu_{p}}{\mu}}\right\} _{\phi_{1},\cdots,\phi_{p}}}\left(\frac{A}{\beta}\right)\left\{ \frac{\Gamma(k_{1}-\mu)\cdots\Gamma(k_{p}-\mu)\ \Gamma(p+\frac{\mu_{0}-m}{\mu})}{\mu\ \Gamma{(k_{1}+\cdots k_{p}-m)}}\left(-\frac{A}{\beta}\Gamma\left(-\mu\right)\right)^{\frac{m}{\mu}-p-\frac{\mu_{0}}{\mu}}\left(-\Gamma(-\mu_{0})\right)\right\} _{\phi_{0}}\right]\nonumber \\
 & + & {\cal O}(\epsilon^{2})\ .\label{pkp_phi}
\end{eqnarray}
The expression for $\langle Z^{m}\rangle$ turns out to be a special
case ($p=0$) of (\ref{pkp_phi}) and therefore at order $\epsilon$
\begin{eqnarray*}
\langle P_{k_{1},\cdots k_{p}}\rangle_{m} & = & \left(\frac{A}{\beta}\right)^{p}\Gamma(-m)\left[\left\{ \frac{\Gamma(k_{1}-\mu_{1})\cdots\Gamma(k_{p}-\mu_{p})\ \ \Gamma(\frac{\mu_{1}+\cdots\mu_{p}-m}{\mu})}{\Gamma(-\frac{m}{\mu})\ \Gamma{(k_{1}+\cdots k_{p}-m)}}\left(-\frac{A}{\beta}\Gamma\left(-\mu\right)\right)^{-\frac{\mu_{1}+\cdots\mu_{p}}{\mu}}\right\} _{\phi_{1},\cdots,\phi_{p}}\right.\\
 & + & \left(\frac{A}{\beta}\right)\frac{\Gamma(k_{1}-\mu)\cdots\Gamma(k_{p}-\mu)\ \ \Gamma(p+1-\frac{m}{\mu})}{\Gamma(-\frac{m}{\mu})\ \Gamma{(k_{1}+\cdots k_{p}-m)}}\left(-\frac{A}{\beta}\Gamma\left(-\mu\right)\right)^{-p-1}\left(-\Gamma(-\mu)\right)\\
 & - & \left(\frac{A}{\beta}\right)\left\{ \frac{\Gamma(k_{1}-\mu)\cdots\Gamma(k_{p}-\mu)\ \ \Gamma(p+\frac{\mu_{0}-m}{\mu})}{\Gamma(-\frac{m}{\mu})\ \Gamma{(k_{1}+\cdots k_{p}-m)}}\left(-\frac{A}{\beta}\Gamma\left(-\mu\right)\right)^{-p-\frac{\mu_{0}}{\mu}}\left(-\Gamma(-\mu_{0})\right)\right\} _{\phi_{0}}\\
 & + & \left(\frac{A}{\beta}\right)\left\{ \frac{\Gamma(k_{1}-\mu)\cdots\Gamma(k_{p}-\mu)\ \ \Gamma(p-\frac{m}{\mu})\ \Gamma(\frac{\mu_{0}-m}{\mu})}{\Gamma(-\frac{m}{\mu})^{2}\ \Gamma{(k_{1}+\cdots k_{p}-m)}}\left(-\frac{A}{\beta}\Gamma\left(-\mu\right)\right)^{-p-\frac{\mu_{0}}{\mu}}\left(-\Gamma(-\mu_{0})\right)\right\} _{\phi_{0}}\\
 & - & \left.\vphantom{\left\{ \left(-\frac{A}{\beta}\Gamma\left(-\mu\right)\right)^{-\frac{\mu_{1}+\cdots\mu_{p}}{\mu}}\right\} _{\phi_{1},\cdots,\phi_{p}}}\left(\frac{A}{\beta}\right)\frac{\Gamma(k_{1}-\mu)\cdots\Gamma(k_{p}-\mu)\ \ \Gamma(p-\frac{m}{\mu})\ \Gamma(1-\frac{m}{\mu})}{\Gamma(-\frac{m}{\mu})^{2}\ \Gamma{(k_{1}+\cdots k_{p}-m)}}\left(-\frac{A}{\beta}\Gamma\left(-\mu\right)\right)^{-p-1}\left(-\Gamma(-\mu)\right)\right]\\
 & + & O(\epsilon^{2})\ .
\end{eqnarray*}
After a (tedious but) straightforward calculation where we have used
two simple properties ($\Gamma'(z+1)=z\Gamma'(z)+\Gamma(z)$ and $\Gamma''(z+1)=z\Gamma''(z)+2\Gamma'(z)$)
of Gamma functions one gets 
\begin{eqnarray}
\langle P_{k_{1},k_{2},...k_{p}}\rangle_{m}=(-)^{p}\frac{\Gamma(p-\frac{m}{\mu})}{\Gamma(-\frac{m}{\mu})}\ \frac{\Gamma(-m)}{\Gamma(k_{1}+\cdots k_{p}-m)}\ \frac{\Gamma(k_{1}-\mu)}{\Gamma(-\mu)}\cdots\frac{\Gamma(k_{p}-\mu)}{\Gamma(-\mu)}\nonumber \\
\times\left[1+2\frac{\epsilon}{\beta^{2}}\log\left(-\frac{A}{\beta}\Gamma\left(-\mu\right)\right)\left(-\frac{\Sigma_{1}}{\mu}+\frac{p}{\mu}\frac{\Gamma'(-{\mu})}{\Gamma(-\mu)}-\frac{m}{\mu^{3}}\frac{\Gamma'(-\frac{m}{\mu})}{\Gamma(-\frac{m}{\mu})}+\frac{m}{\mu^{3}}\frac{\Gamma'(p-\frac{m}{\mu})}{\Gamma(p-\frac{m}{\mu})}\right)\right.\nonumber \\
+\frac{\epsilon}{\beta^{2}}\left(-{\Sigma_{2}}+2\frac{\Sigma_{1}}{\mu}\frac{\Gamma'(p-\frac{m}{\mu})}{\Gamma(p-\frac{m}{\mu})}-{\frac{2m\Gamma'(-{\mu})\,\Gamma'(-\frac{m}{\mu})}{\mu^{2}\Gamma(-{\mu})\,\Gamma(-\frac{m}{\mu})}+\frac{2m\Gamma'(-{\mu})\,\Gamma'(p-\frac{m}{\mu})}{\mu^{2}\Gamma(-{\mu})\,\Gamma(p-\frac{m}{\mu})}}\right.\nonumber \\
\left.\left.-\frac{2p\Gamma'(-{\mu})\,\Gamma'(p-\frac{m}{\mu})}{\mu\Gamma(-{\mu})\,\Gamma(p-\frac{m}{\mu})}+p\frac{\Gamma''(-{\mu})}{\Gamma(-{\mu})}+\frac{m\Gamma''(-\frac{m}{\mu})}{\mu^{3}\Gamma(-\frac{m}{\mu})}-\frac{2\Gamma'(-\frac{m}{\mu})}{\mu^{2}\Gamma(-\frac{m}{\mu})}-\frac{m\Gamma''(p-\frac{m}{\mu})}{\mu^{3}\Gamma(p-\frac{m}{\mu})}+\frac{2\Gamma'(p-\frac{m}{\mu})}{\mu^{2}\Gamma(p-\frac{m}{\mu})}\right)\right]\nonumber \\
+{\cal O}(\epsilon^{2})\label{XXX}
\end{eqnarray}
where 
\[
\Sigma_{1}=\sum_{i=1}^{p}\frac{\Gamma'(k_{i}-\mu)}{\Gamma(k_{i}-\mu)}\ \ \ \ \ ;\ \ \ \ \ \ \Sigma_{2}=\sum_{i=1}^{p}\frac{\Gamma''(k_{i}-\mu)}{\Gamma(k_{i}-\mu)}
\]
We checked that for $p=1$ this formula reduces to (\ref{R1}) in
the limit $m\to0$.

Using expressions (\ref{P2-infinite},\ref{epsilon-def}) and (\ref{eq:A_dfnn})
for $\mu,\epsilon$ and $A$ gives the $1/N$ corrections in terms
of $T$ and $T_{c}$.

\subsection[The non-integer moments of the partition
function]{The non-integer moments $\langle Z^{m}\rangle$ of the partition
function}

A by-product of the above calculation (obtained by setting $p=0$
in (\ref{pkp_phi})), is the expression of the non-integer moments
$\langle Z^{m}\rangle$ for $0<m<1$. At leading order in $\epsilon$
it gives 
\begin{equation}
\left\langle Z^{m}\right\rangle =e^{-\beta mE_{0}}\frac{\Gamma(-\frac{m}{\mu})}{\mu\ \Gamma(-m)}\left(-\frac{A}{\beta}\Gamma\left(-\mu\right)\right)^{\frac{m}{\mu}}+{\cal O}(\epsilon)\ .\label{zm}
\end{equation}
Then replacing $A$ by its expression (\ref{eq:A_dfnn}) 
\begin{equation}
{\langle Z^{m}\rangle_{{\rm exact}}\simeq\frac{1}{\mu}\frac{\Gamma\left(-\frac{m}{\mu}\right)}{\Gamma(-m)}(N\pi\beta^{2}J^{2})^{-\frac{m}{2\mu}}\left(-\Gamma\left(-{\mu}\right)\right)^{\frac{m}{\mu}}\ e^{N\beta mJ\sqrt{\log2}}}+{\cal O}(\epsilon)\ .\label{zm1}
\end{equation}
This expression is obtained under the condition (\ref{condition})
for $0<m<\mu=T/T_{c}<1$. In the limt $m\to0$ one recovers the free
energy \citep{derrida_random-energy_1980,derrida_random-energy_1981,galves_fluctuations_1989,cook_finite-size_1991}.
For $m>0$, the $N$ dependence is also the same as in \citep{gardner_probability_1989}.
If the condition $0<m<\mu=T/T_{c}<1$ is not satisfied then on would
need to expand $\rho(E)$ around an energy different from $E_{0}$
(see (\ref{eq:rho_in_epsilon_form})). For example, for $\mu>1$,
that is in the high temperature phase, the configurations which contribute
most are those with an energy $\simeq-NJ^{2}/2/T$ (see \citep{derrida_random-energy_1981})
and one could in principle repeat the above calculation (done for
$\mu=T/T_{c}<1$) by starting with the approximation (\ref{eq:rho_in_epsilon_form0})
with $E_{0}=-NJ^{2}/2/T$.

\section{The replica approach for the overlaps}
\label{replicas}
In this section we are going to see that expression (\ref{XXX}) obtained
by a direct calculation {\red without use of replicas} is fully consistent with a broken symmetry
of replicas when one lets the number of blocks and the sizes of the
blocks fluctuate (with negative variances).

\subsection{The Parisi ansatz}

In the Parisi replica approach \citep{parisi_sequence_1980,parisi_order_1980,mezard_spin_1987}
to compute $\langle Z^{m}\rangle$, the symmetry between the $m$
replicas is broken, meaning that the $m$ replicas are grouped into
blocks. For example, at the level of a single step of symmetry breaking,
(for the REM it is well known that a single step is sufficient {\red \citep{gross_simplest_1984,mezard_information_2009,biroli_random_2012}})
this means that $\langle Z^{m}\rangle$ is dominated by situations
where the $m$ replicas are grouped into $r$ blocks of $\mu$ replicas.
Then the weighted overlaps are given by 
\begin{equation}
\langle P_{k_{1},\cdots k_{p}}\rangle_{m}=\left(\frac{r!}{(r-p)!}\right)\left(\prod_{i=1}^{p}\frac{\mu!}{(\mu-k_{i})!}\right)\left(\frac{(m-k_{1}-\cdots k_{p})!}{m!}\right)\ .\label{replic0}
\end{equation}
Expression (\ref{replic0}) as well as its generalization (\ref{replic1})
will be established in section \ref{letting}. In short, the first
factor in (\ref{replic0}) counts the number of ways of choosing $p$
blocks among the $r$ blocks, the product counts the number of ways
of choosing $k_{1},\cdots k_{p}$ replica in each of the $p$ blocks
of $\mu$ replica each, the last term is the normalization which corresponds
to the number of ways of choosing $k_{1}+\cdots k_{p}$ replicas among
$m$.

When $p,k_{1},k_{2},\cdots k_{p}$ are integers, (\ref{replic0})
is a rational function of the parameters $m$, $r$ and $\mu$. Therefore
it can be analytically continued to non-integer values of these parameters
$m$, $r$ and $\mu$ and coincides with the following rational function
\begin{equation}
\langle P_{k_{1},\cdots k_{p}}\rangle_{m}=\left(\frac{\Gamma(p-r)}{-\Gamma(-r)}\right)\left(\prod_{i=1}^{p}\frac{\Gamma(k_{i}-{\mu})}{-\Gamma(-{\mu})}\right)\left(\frac{-\Gamma(-m)}{\Gamma(k_{1}+\cdots k_{p}-m)}\right)\label{replic1}
\end{equation}
If all blocks have the same size $\mu$, the number of blocks is obviously
\[
r=\frac{m}{\mu}
\]
and with this choice, one can see that (\ref{replic1}) reduces to
the leading order of (\ref{XXX}). So the broken replica symmetry
does give the correct expression for the large $N$ limit of the overlaps.

\subsection{Letting the number of blocks and their sizes fluctuate}

\label{letting} Now we want to let the number $r$ of blocks, and
the numbers $\mu_{1},\cdots\mu_{k}$ of replicas in these blocks fluctuate.
Then (\ref{replic1}) becomes 
\begin{equation}
\langle P_{k_{1},\cdots k_{p}}\rangle_{m}=\left\langle \ \left(\frac{\Gamma(p-r)}{-\Gamma(-r)}\right)\left(\prod_{i=1}^{p}\frac{\Gamma(k_{i}-\mu_{i})}{-\Gamma(-\mu_{i})}\right)\left(\frac{-\Gamma(-m)}{\Gamma(k_{1}+\cdots k_{p}-m)}\right)\ \right\rangle _{r,\mu_{1},\cdots\mu_{p}}.\label{replic2}
\end{equation}
\textit{Derivation of (\ref{replic2}):} Let us now explain how (\ref{replic2})
can be derived. For the Poisson REM one can write the following exact
expression of $\langle Z^{m}\rangle$ when $m$ is an integer,

\begin{equation}
\langle Z^{m}\rangle=\sum_{r\ge1}\frac{m!}{r!}\ \sum_{\mu_{1}\ge1}\cdots\sum_{\mu_{r}\ge1}\frac{\Psi(\mu_{1})}{\mu_{1}!}\cdots\frac{\Psi(\mu_{r})}{\mu_{r}!}\ \delta[m,\mu_{1}+\cdots\mu_{r}]\label{formula0}
\end{equation}
where 
\[
\Psi(\mu)=\int\rho(E)e^{-\mu\beta E}dE
\]
One can evaluate in the same way, still for integer $m$, 
\begin{equation}
\langle Z^{m}P_{k_{1},\cdots{k_{p}}}\rangle=\sum_{r\ge1}\frac{(m-k_{1}-\cdots k_{p})!}{(r-p)!}\sum_{\mu_{1}\ge1}\cdots\sum_{\mu_{r}\ge1}\frac{\Psi(\mu_{1})}{(\mu_{1}-k_{1})!}\cdots\frac{\Psi(\mu_{p})}{(\mu_{p}-k_{p})!}\ \frac{\Psi(\mu_{p+1})}{\mu_{p+1}!}\cdots\frac{\Psi(\mu_{r})}{\mu_{r}!}\ \delta[m,\mu_{1}+\cdots\mu_{r}]\label{formula1}
\end{equation}

Here the convention is $(-n)!=\infty$ for $n=1,2\cdots$. Taking
the ratio of (\ref{formula1}) and (\ref{formula0}) one gets 
\begin{eqnarray}
\langle P_{k_{1},\cdots{k_{p}}}\rangle_{m}\equiv\frac{\langle Z^{m}P_{k_{1},\cdots{k_{p}}}\rangle}{\langle Z^{m}\rangle}=\left\langle \frac{(m-k_{1}-\cdots k_{p})!}{m!}\left(\prod_{i=1}^{p}\frac{\mu_{i}!}{(\mu_{i}-k_{i})!}\right)\frac{r!}{(r-p)!}\right\rangle _{r,\mu_{1},\cdots\mu_{p}}\label{replic3}
\end{eqnarray}
where $\langle.\rangle_{r,\mu_{1},\cdots\mu_{p}}$ simply means an
average over $r,\mu_{1},\cdots\mu_{p}$. That is, for a test function
$G$, 
\[
\langle G(r,\mu_{1},\cdots\mu_{p})\rangle_{r,\mu_{1},\cdots\mu_{p}}=\frac{1}{\langle Z^{m}\rangle}\sum_{r\ge1}\frac{m!}{r!}\ \sum_{\mu_{1}\ge1}\cdots\sum_{\mu_{r}\ge1}\frac{\Psi(\mu_{1})}{\mu_{1}!}\cdots\frac{\Psi(\mu_{r})}{\mu_{r}!}\delta[m,\mu_{1}+\cdots\mu_{r}]G(r,\mu_{1},\cdots\mu_{p})\ .
\]

Expression (\ref{replic3}) is exact for any positive integer $m$.
It has been derived when all the parameters are integers. For fixed
integer values of $p$, $k_{1},\cdots k_{p}$, it is a rational function
of the parameters $m$, $\mu_{1},\cdots\mu_{p}$. Therefore it can
be analytically continued to non integer values of these parameters
and it coincides with (\ref{replic2}). This completes our derivation
of (\ref{replic0},\ref{replic1},\ref{replic2}).

1
\subsection{Characteristics of the fluctuations}

We now try to see in (\ref{replic2}) what kind of fluctuations of
the number $r$ of blocks and the numbers $\mu_{1}\cdots\mu_{k}$
of replicas in each block would enable us to recover the finite size
corrections obtained in (\ref{XXX}) by a direct calculation.

We have checked, by \textquotedbl{}a tedious but straightforward calculation\textquotedbl{}
that if we write for $1\le i\le k$ 
\begin{eqnarray}
\mu_{i}=\mu+\psi_{i}\label{fluctuations}\\
r=\frac{m}{\mu}+\rho\nonumber 
\end{eqnarray}
 (\ref{replic2}) becomes equivalent to (\ref{XXX}) provided that
\begin{eqnarray}
\langle\psi_{i}\rangle & = & \epsilon\left[\frac{2}{\beta^{2}\mu}\log\left(-\frac{A}{\beta}\Gamma(-\mu)\right)+\frac{2}{\beta^{2}}\frac{\Gamma'(-\mu)}{\Gamma(-\mu)}-\frac{2}{\beta^{2}\mu}\frac{\Gamma'(-\frac{m}{\mu})}{\Gamma(-\frac{m}{\mu})}\right]\nonumber \\
\langle\rho\rangle & = & -\frac{m}{\mu^{2}}\langle\psi_{i}\rangle-\frac{2}{\mu^{2}\beta^{2}}\epsilon\nonumber \\
\langle\psi_{i}^{2}\rangle & = & -\frac{2}{\beta^{2}}\epsilon\label{fluct0}\\
\langle\psi_{i}\psi_{j}\rangle & = & 0\nonumber \\
\langle\rho\psi_{i}\rangle & = & \frac{2}{\mu\beta^{2}}\epsilon\nonumber \\
\langle\rho^{2}\rangle & = & -\frac{2m}{\mu^{3}\beta^{2}}\epsilon\ .\nonumber 
\end{eqnarray}

In terms of $\mu=T/T_{c},\beta=1/T$ and $N$ these expressions become
\begin{eqnarray}
\langle\psi_{i}\rangle & = & \frac{2}{N\beta^{2}J^{2}\mu}\left[\log\left(-\frac{\Gamma(-\mu)}{\beta J\sqrt{N\pi}}\right)+\frac{\mu\Gamma'(-\mu)}{\Gamma(-\mu)}-\frac{\Gamma'(-\frac{m}{\mu})}{\Gamma(-\frac{m}{\mu})}\right]\nonumber \\
\langle\rho\rangle & = & -\frac{m}{\mu^{2}}\langle\psi_{i}\rangle-\frac{2}{N\mu^{2}\beta^{2}J^{2}}\nonumber \\
\langle\psi_{i}^{2}\rangle & = & -\frac{2}{N\beta^{2}J^{2}}\label{fluctu}\\
\langle\psi_{i}\psi_{j}\rangle & = & 0\nonumber \\
\langle\rho\psi_{i}\rangle & = & \frac{2}{N\mu\;\beta^{2}J^{2}}\nonumber \\
\langle\rho^{2}\rangle & = & -\frac{2m}{N\mu^{3}\beta^{2}J^{2}}\ .\nonumber 
\end{eqnarray}
So the $1/N$ corrections we calculated directly in (\ref{XXX}) can
indeed be interpreted as fluctuations of the number $r$ of blocks
and of the sizes $\mu_{i}$ of the blocks in Parisi's ansatz. The
only price we pay is to allow negative variances.

\section{{\red The exact  non-integer moments of  the partition function written  in terms of "complex replica numbers"}}

{\red In this section we write  the exact  expression (\ref{zm1}) of $\langle Z^{m}\rangle$ 
 for the non-integer moments  of the partition function (which was obtained without using replicas)  
  in terms of  contour integrals over what can be thought of as 
  "complex replica numbers". We  then show that one step replica symmetry breaking arises naturally from the saddle point of these integrals without
  making  the Parisi ansatz. We also find that, without any additional assumptions, 
 the fluctuations of the "complex replica numbers"  in the imaginary direction correspond to the negative 
variances   (\ref{fluctu}) we observed in the replica calculation of section \ref{replicas} .}

\subsection[The contour integral representation of the non-integer moments]{The {\red contour integral representation of the} non-integer moments $\langle Z^{m}\rangle$
for $0<m<1$ }

Our starting point is the following representation, valid for $0<m<1$,
of the non integer moments 
\begin{equation}
\langle Z^{m}\rangle=\frac{1}{\Gamma(-m)}\int_{0}^{\infty}{\rm d}t\ t^{-m-1}\left(\left\langle e^{-tZ}\right\rangle -1\right)\ .\label{Za}
\end{equation}

We have seen (\ref{Znu0},\ref{Znu}) that 
\begin{equation}
\langle e^{-tZ}\rangle=\exp\left[\int\rho(E)dE\left(\exp[-te^{-\beta E}]-1\right)\right]\ .\label{Zb}
\end{equation}

\begin{figure}
\noindent \begin{centering}
\includegraphics{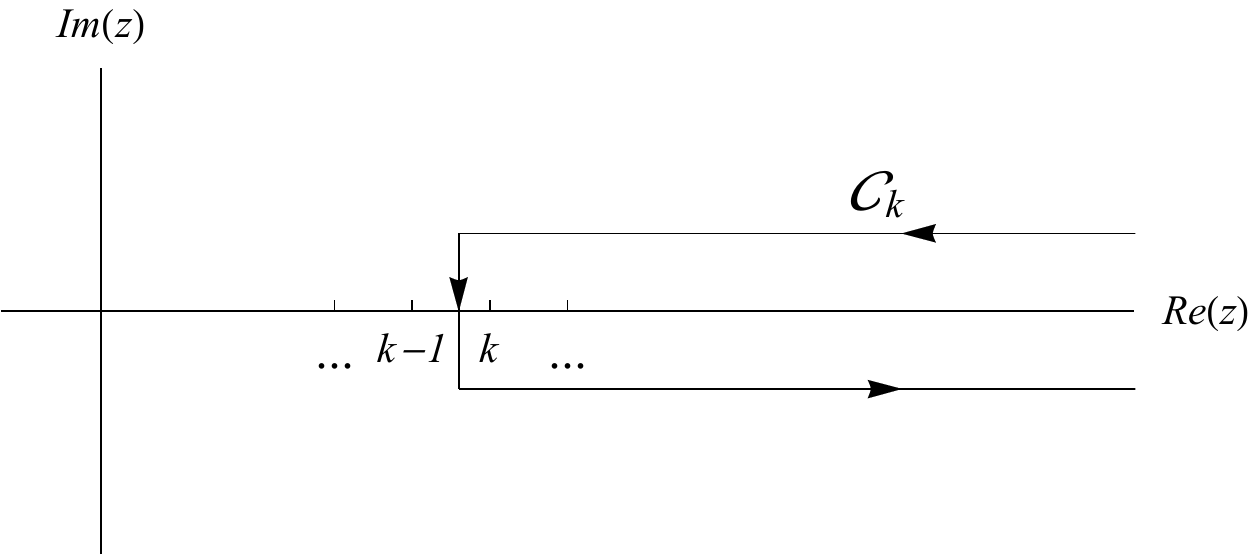}
\par\end{centering}

\protect\caption{The contour ${\cal C}_{k}$ starts at $+\infty+i0$ and ends at $+\infty-i0$
and crosses the real axis between $k-1$ and $k$.\label{fig:Contour}}
\end{figure}

We now need to use the identity 
\begin{equation}
\sum_{p\ge k}f(p)\frac{(-1)^{p}}{p!}=-\int_{{\cal C}_{k}}\frac{dz}{2\pi i}\ \Gamma(-z)f(z)\label{magic}
\end{equation}
 where the contour ${\cal C}_{k}$ starts at $+\infty+i0$ and ends
at $+\infty-i0$ and crosses the real axis between $k-1$ and $k$
(see figure \ref{fig:Contour}). 

This identity is valid for any analytic function $f(z)$ such that
the sum and the integral in (\ref{magic}) converge. Using (\ref{magic})
in (\ref{Za},\ref{Zb}) (see the discussion on convergence in section
\ref{conv} below) one can see that 
\[
\langle Z^{m}\rangle=\frac{-1}{\Gamma(-m)}\int_{0}^{\infty}{\rm dt}\ t^{-m-1}\int_{{\cal C}_{1}}\frac{dr}{2\pi i}\ \Gamma(-r)\left(-\int\rho(E)dE\left(\exp[-te^{-\beta E}]-1\right)\right)^{r}
\]
Using again the identity (\ref{magic}) one gets 
\begin{equation}
\langle Z^{m}\rangle=-\frac{1}{\Gamma(-m)}\int_{0}^{\infty}{\rm d}t\ t^{-m-1}\int_{{\cal C}_{1}}\frac{{\rm d}r}{2\pi i}\ \Gamma(-r)\left(\int_{{\cal C}_{1}}\frac{{\rm d}\mu}{2\pi i}\ \Gamma(-\mu)\ t^{\mu}\int dE\rho(E)e^{-\beta\mu E}\right)^{r}\ .\label{XXX1}
\end{equation}

Let us assume that 
\begin{equation}
\int\rho(E)dE\ e^{-\beta\mu E}=B\exp[N\phi(\mu)]\ .\label{phi-def}
\end{equation}
For example, for the Poisson REM this gives (\ref{eq:energy_dist})
\begin{equation}
B=1\ \ \ \ \ ;\ \ \ \ \phi(\mu)=\log2+\frac{\beta^{2}J^{2}\mu^{2}}{4}\ .\label{phi-rem}
\end{equation}
Making the change of variables $t=e^{Nx}$ equation (\ref{XXX1})
becomes 
\begin{equation}
\langle Z^{m}\rangle=-\frac{N}{\Gamma(-m)}\int_{-\infty}^{\infty}{\rm d}x\ e^{-Nxm}\int_{{\cal C}_{1}}\frac{{\rm d}r}{2\pi i}\ \Gamma(-r)\ \left(B\ \int_{{\cal C}_{1}}\frac{{\rm d}\mu}{2\pi i}\ \Gamma(-\mu)\ e^{N(x\mu+\phi(\mu))}\right)^{r}\ .\label{XXX2}
\end{equation}
This expression is exact. Our goal now is to get its large $N$ behaviour.
To do so we found it more convenient to perform the saddle point calculation
in the following order: first the integral over $\mu$, then the integral
over $x$, then the integral over $r$.

For large $N$, we evaluate the integral over $\mu$ using a saddle
point, at a value of $0<\mu<1$ on the real axis to give 
\begin{equation}
\langle Z^{m}\rangle=-\frac{-N}{\Gamma(-m)}\int_{-\infty}^{\infty}{\rm d}x\ e^{-Nmx}\int_{{\cal C}_{1}}\frac{{\rm d}r}{2\pi i}\ \Gamma(-r)\left(\frac{-\Gamma(-\mu)\ B}{\sqrt{2\pi N\phi''(\mu)}}\right)^{r}\ e^{N(x\mu+\phi(\mu))r}\label{XXX2a}
\end{equation}
where the saddle point value $\mu$ has become a function of $x$
and is solution of 
\begin{equation}
x+\phi'(\mu)=0\ .\label{mu-saddle}
\end{equation}
\\
 Now that $\mu$ is a function of $x$, one can calculate in (\ref{XXX2a})
the saddle point with respect to $x$ which is determined { (together
with (\ref{mu-saddle})) by} 
\begin{equation}
r\mu-m=0\ .\label{x-saddle}
\end{equation}
Equations (\ref{mu-saddle}) and (\ref{x-saddle}) give the saddle
point values $\mu$ and $x$ in terms of $r$ 
\begin{equation}
\mu=\frac{m}{r}\ \ \ \ ;\ \ \ \ x=-\phi'\left(\frac{m}{r}\right)\label{mux}
\end{equation}
so that (\ref{XXX2a}) becomes 
\begin{equation}
\langle Z^{m}\rangle\simeq\frac{-1}{\ \Gamma(-m)}\int_{{\cal C}_{1}}\frac{{\rm d}r}{i\sqrt{2\pi}}\ \Gamma(-r)\frac{\sqrt{N\phi''(\mu)}}{\sqrt{r}}\left(\frac{-\Gamma(-\mu)\ B}{\sqrt{2\pi N\phi''(\mu)}}\right)^{r}\ e^{Nr\phi(\frac{m}{r})}\label{XXX3}
\end{equation}
where we have used (see (\ref{mu-saddle})) that $1+\phi''(\mu)\frac{d\mu}{dx}=0$.
\\
 Now looking for the saddle point in $r$ one gets that it should
satisfy 
\begin{equation}
\phi\left(\frac{m}{r}\right)-\frac{m}{r}\phi'\left(\frac{m}{r}\right)=0\ \ \ \ {\rm i.e.}\ \ \ \ \phi(\mu)-\mu\phi'(\mu)=0\label{r-saddle}
\end{equation}
and (\ref{XXX3}) becomes 
\begin{equation}
\langle Z^{m}\rangle\simeq\frac{r}{m}\ \frac{\Gamma(-r)}{\Gamma(-m)}\ \left(\frac{-\Gamma(-\mu)\ B}{\sqrt{2\pi N\phi''(\mu)}}\right)^{r}\ e^{Nr\phi(\frac{m}{r})}=\frac{1}{\mu}\ \frac{\Gamma(-\frac{m}{\mu})}{\Gamma(-m)}\ B^{\frac{m}{\mu}}\left(\frac{-\Gamma(-\mu)}{\sqrt{2\pi N\phi''(\mu)}}\right)^{\frac{m}{\mu}}e^{N\frac{m}{\mu}\phi(\mu)}\ .\label{XXX4}
\end{equation}
{\red This is our result,
when $0<m<\mu<1$, 
 for the non integer moments $\langle Z^{m}\rangle$  with $\phi$ given by(\ref{phi-def}) or (\ref{phi-rem})  and $\mu$  the solution of (\ref{r-saddle}). 
 It  was obtained without using the
Parisi's symmetry breaking scheme. However, as we discuss further below, the correspondence
 between the $m$, $r$ and $\mu$ integration variables used here and the replica numbers used in section \ref{replicas} (see for example formula \ref{formula0})  is  rather compelling. For this reason we will sometimes refer to the $m$, $r$ and $\mu$ used in this section as "complex
 replica numbers". }

\subsection{Some remarks on the replica calculation}

At this point we would like to make some remarks on the significance
of the results from this approach to the replica calculation:

\  \ \\
 \textit{Remark 1}: The above saddle point estimate is legitimate
only if the saddle point value of $r$ is between $0$ and $1$ (i.e.
in the range where the contour ${\cal C}_{1}$ crosses the real axis).
If not one can deform the contour to pass through the saddle point
but one should not forget the contribution of the poles at the integer
values of $r$. This is what happens in particular in the high temperature
phase {\red \cite{campellone_non-perturbative_1995,meiners_accuracy_2013}}.

\ \\
 \textit{Remark 2 }: A similar calculation could be done for $m>1$.
The difference would be to replace the integration contour ${\cal C}_{1}$
of $r$ in (\ref{XXX1}) by ${\cal C}_{k}$ for $k-1<m<k$. The rest
of the calculation would be very similar. 

\ \\
{\red 
 \textit{Remark 3 :} A formula similar to (\ref{magic})
was already used in \citep{campellone_replica_2009} (see also \citep{ogure_exact_2004}).
Here    we use it twice starting from an exact expression  
and the one step replica symmetry breaking   structure arises naturally as the saddle point of an exact expression  (\ref{XXX2}) in  the large $N$ limit. The finite size corrections follow from fluctuations in the "complex replica numbers" about this  saddle point.

\ \\
 \textit{Remark 4 :} 
Several works in the past 
\cite{nieuwenhuizen_puzzle_1996,ferrero_fluctuations_1996,ferrero_infrared_1996}
have discussed  the necessity of calculating 
the fluctuations around  the saddle point to obtain the leading  finite size corrections, using in particular linear response theory to determine the fluctuations of the Parisi function. The main difference  between these works and   our present approach is that our starting point  to calculate finite size corrections is not based on Parisi's ansatz.    It would be nice to see whether our results (\ref{fluctu}) could be recovered as the large $p$ limit in the replica calculations  \cite{nieuwenhuizen_puzzle_1996,ferrero_fluctuations_1996} of the $p$ spin models.

\ \\
 \textit{Remark  5} :} For the REM (\ref{phi-rem}) the saddle point
equations (\ref{r-saddle}) gives for $\mu=2\sqrt{\log2}/\beta J$
which agrees with the definition of $\mu$ in (\ref{P2-infinite}).
Then as $B=1$, one can check that (\ref{XXX4}) does coincide with
(\ref{zm1}). So the {\red exact "complex replica"} approach based on (\ref{XXX2}) indeed
agrees with the direct calculation leading to (\ref{zm1}). Note that
to get the right prefactor in (\ref{XXX2}) it was necessary to integrate
over the fluctuations of the parameters $r,\mu$ and $x$ meaning
that we had to include the fluctuations around Parisi's ansatz.

\subsection{How the replica numbers became complex}

There is a remarkable similarity between the expressions (which are
both exact) (\ref{formula0}) and (\ref{XXX1}) of $\langle Z^{m}\rangle$
for integer and non-integer $m$: in (\ref{XXX1}) the number $r$
of blocks and the sizes $\mu_{1},\cdots\mu_{r}$ of the blocks are
not integer anymore (they have even become complex!). Going from integer
$m$ to non-integer $m$, one has to replace the measure (\ref{formula0})
\[
\sum_{r\ge1}\frac{m!}{r!}\ \sum_{\mu_{1}\ge1}\cdots\sum_{\mu_{r}\ge1}\frac{\delta[m,\mu_{1}+\cdots\mu_{r}]}{\mu_{1}!\cdots\mu_{r}!}
\]
by (\ref{XXX1}) 
\[
\frac{-1}{2\pi i\Gamma(-m)}\int_{0}^{\infty}{\rm d}t\ t^{-m-1}\int_{{\cal C}_{1}}{\rm d}r\ \Gamma(-r)\left(\frac{1}{2\pi i}\right)^{r}\int_{{\cal C}_{1}}{\rm d}\mu_{1}\ \Gamma(-\mu_{1})\cdots\int_{{\cal C}_{1}}{\rm d}\mu_{r}\ \Gamma(-\mu_{r})\ t^{\mu_{1}+\cdots\mu_{r}}\ .
\]
These two measures can be rewritten respectively as 
\begin{equation}
\int_{-\infty}^{\infty}\frac{dx}{2\pi}e^{-mix}\sum_{r\ge1}\frac{m!}{r!}\ \sum_{\mu_{1}\ge1}\cdots\sum_{\mu_{r}\ge1}\frac{e^{ix(\mu_{1}+\cdots\mu_{r})}}{\mu_{1}!\cdots\mu_{r}!}\label{discrete}
\end{equation}
and 
\begin{equation}
-\int_{-\infty}^{\infty}\frac{{\rm d}x}{2\pi i}\ e^{-mx}\int_{{\cal C}_{1}}{\rm d}r\ \frac{\Gamma(-r)}{\Gamma(-m)}\int_{{\cal C}_{1}}{\rm d}\mu_{1}\ \frac{\Gamma(-\mu_{1})}{2\pi i}\cdots\int_{{\cal C}_{1}}{\rm d}\mu_{r}\ \frac{\Gamma(-\mu_{r})}{2\pi i}\ e^{x(\mu_{1}+\cdots\mu_{r})}\ .\label{continuous}
\end{equation}
Comparing these two expressions, we see that, up to factors $i$ or
$-1$, the sums have become integrals, the integers $r,\mu_{1},\mu_{p}$
have become complex, and the inverse factorials $1/n!$   have been
replaced by $\frac{\Gamma(-n)}{2\pi i}$.
 For the REM, this may shed some light on the mystery of Parisi's theory.  

\subsection{Why the diverging integrals or series are harmless}

\label{conv} In the last step to obtain (\ref{XXX1}) we wrote 
\begin{equation}
\int\rho(E)dE\left(\exp[-te^{-\beta E}]-1\right)=\int_{{\cal C}_{1}}\frac{{\rm d}\mu}{2\pi i}\ \Gamma(-\mu)\ t^{\mu}\int dE\rho(E)e^{-\beta\mu E}\label{harmless1}
\end{equation}
which in the case of the REM (\ref{phi-rem}) gives

\begin{equation}
\int\rho(E)dE\left(\exp[-te^{-\beta E}]-1\right)=\int_{{\cal C}_{1}}\frac{{\rm d}\mu}{2\pi i}\ \Gamma(-\mu)\ t^{\mu}\;2^{N}\;\exp\left[\mu^{2}\frac{N\beta^{2}J^{2}}{4}\right]\ .\label{harmless2}
\end{equation}
The integral in the r.h.s. of (\ref{harmless2}) clearly diverges
as $\mu\to\infty\pm i0$.

The same would be true, in the limit $\mu\to\infty$, for the following
power series expansion 
\begin{equation}
\int\rho(E)dE\left(\exp[-te^{-\beta E}]-1\right)=\sum_{\mu\ge1}\ \frac{(-t)^{\mu}}{\mu!}2^{N}\exp\left[\mu^{2}\frac{N\beta^{2}J^{2}}{4}\right]\label{harmless3}
\end{equation}
We think that these divergences are harmless for the following reason:
we know that in the low temperature phase of the REM, everything is dominated
by the energies close to $E_{0}$ given by (\ref{E0-def}). Therefore
the $1/N$ corrections of the present paper would remain unchanged
if one would replace the density $\rho(E)$ by a density $\tilde{\rho}(E)$
which is identical to $\rho(E)$ in the neighborghood of $E_{0}$.
For example we could choose 
\[
\tilde{\rho}(E)=\left\{ \begin{array}{ccc}
\rho(E) & {\rm if} & |E|<2|E_{0}|\\
0 &  & |E|>2|E_{0}|
\end{array}\right.\ .
\]
With the distribution $\tilde{\rho}(E)$ the above integral and sum
would become convergent while none of our results would be modified.

Diverging series appear frequently in replica calculations \citep{dotsenko_bethe_2010,calabrese_exact_2011,dotsenko_one_2011}
in particular in the context of the KPZ equation and it would of course
be interesting to see whether a similar reason could be invoked there
to justify the manipulation of diverging series or integrals.

\subsection{The fluctuations close to the saddle point}

\label{saddlep} In the derivation of (\ref{XXX4}), we started from
the exact expression (\ref{XXX2}) and we performed three saddle point
calculations. The saddle point values were given by (\ref{mu-saddle},\ref{x-saddle},\ref{r-saddle})
\[
x=-\phi'(\mu)\ \ \ \ ;\ \ \ \ r=\frac{m}{\mu}\ \ \ \ ;\ \ \ \ \phi(\mu)-\mu\phi'(\mu)=0\,.
\]
One can now try to characterize the fluctuations responsible of the
large $N$ corrections at these saddle points.

If we apply the formulas derived in Appendix B to the integral (\ref{XXX3})
over $r$, one gets for the fluctuations (\ref{fluctuations}) near
this saddle point, after replacing $r$ by $m/\mu$ 
\[
\langle\rho\rangle=\frac{1}{N}\left[-\frac{r^{2}}{m^{2}\phi''(\mu)}-\frac{r^{2}}{2m}\frac{\phi'''(\mu)}{\phi''(\mu)^{2}}-\frac{r^{3}}{m^{2}\phi''(\mu)}\left(-\frac{\Gamma'(-\frac{m}{\mu})}{\Gamma(-\frac{m}{\mu})}+\log\left[-\frac{\Gamma(-\mu)\ B}{\sqrt{2\pi N\phi''(\mu)}}\right]\right)+{\mu}\frac{\Gamma'(-\mu)}{\Gamma(-\mu)}\right]
\]
\[
\langle\rho^{2}\rangle=-\frac{1}{N}\frac{m}{\mu^{2}\phi''(\mu)}
\]
which in the case of the REM (\ref{phi-rem}) is fully consistent
with (\ref{fluctu}).

The calculation of the fluctuations of $r$ is easier because in our
saddle point calculation the integral over $r$ was performed last.
The other fluctuations predicted in (\ref{fluctu}) are more difficult
to recover because the saddle point in $\mu$ depends on $x$ which
itself depends on $r$ and that both $x$ and $r$ fluctuate. So the
fluctuations of $\mu$ would combine its own fluctuations with those
induced by the fluctuations of $x$ and $r$. Moreover here there
is a single (or may be $r$) variable $\mu$ while in (\ref{fluctu})
there are $p$ of them. Because of these difficulties we did not calculate
the fluctuations of $\mu$. We however believe that, if done correctly,
the fluctuations of $\mu$ should be consistent with (\ref{fluctu}).

\section{Conclusion}

In the present paper we have developed a systematic  way of computing
finite size effects for the REM. Our approach led to explicit expressions
of the leading corrections to the overlaps (\ref{R2},\ref{XXX})
and of the prefactor of the non-integer moments of the partition function
(\ref{zm1}). We have shown that these results  can be interpreted
as fluctuations {\red with negative variance} (\ref{fluctu}) of the number and size of the blocks in the broken replica symmetry language.

{\red Our exact  expression (\ref{XXX2}) for the non-integer moments
 of the partition function can be written in terms of coupled contour integrals over what can be thought of as   "complex replica numbers". One step replica symmetry breaking then arises naturally from the saddle point of these integrals without the need of using the conventional replica approach or the Parisi
ansatz. We also find that 
 the fluctuations  of the "complex replica numbers"  in the imaginary direction correspond to the negative variances we observed in the replica calculation.}
  
One can try to extend our approach to calculate the fluctuations and
the finite size corrections in a number of other cases such as 
{\red 
the REM with complex temperatures \citep{allakhverdyan_finite_1997,derrida_zeroes_1991,moukarzel_numerical_1991,saakian_random_2000,saakian_phase_2009}},
generalised random energy models, or directed polymers on a tree \citep{cook_finite-size_1991}.
More challenging would be to  see 
{\red    whether
our approach could  give  an altenative to the replica method  \cite{nieuwenhuizen_puzzle_1996,ferrero_fluctuations_1996,ferrero_infrared_1996}
for }
models of disordered systems \citep{hukushima_replica-symmetry-breaking_2000,gardner_spin_1985,dillmann_finite-size_1998}
such as glasses (see \citep{biroli_random_2012} for a recent review)
optimisation problems (see \citep{mezard_information_2009} for references)
which, like the REM, exhibit a one step replicas symmetry breaking (1RSB)
\citep{gross_simplest_1984,bouchaud_universality_1997}.

Lastly, generalizing a formula like (\ref{XXX2}) to some other disordered
systems, starting with the Sherrington Kirpatrick model, would certainly
improve our understanding of the applicability and limitations of
the replica approach.
\begin{description}
\item [{Acknowledgements}]~
\end{description}
We would like to thank the Higgs Centre and the Institute for Condensed
Matter and Complex Systems at the University of Edinburgh for their
kind hospitality and Martin Evans, in particular, for his support
and encouragement.

\section*{Appendices}

\appendix

\section{Difference between the REM and the Poisson REM\label{sec:Poisson_rem}}

In this appendix we show that the difference between the original
REM and the Poisson REM of section \ref{poisson} is exponentially
small in the system size $N$.

In the original REM one considers a system with $2^{N}$ configurations
${\cal C}$ whose energies $E({\cal C)}$ are i.i.d. random variables
distributed according to a Gaussian distribution 
\begin{equation}
P(E)=\frac{1}{\sqrt{\pi NJ^{2}}}\exp\left[-\frac{E^{2}}{NJ^{2}}\right]\label{P(E)}
\end{equation}

and the partition function is 
\[
Z=\sum_{{\cal C}}e^{-\beta E({\cal C)}}\ .
\]
The generating function of the partition function is simply given
by 
\begin{equation}
\left\langle e^{-tZ}\right\rangle _{\text{REM}}=\left[\int e^{-te^{-\beta E}}\ P(E)\ dE\right]^{2^{N}}\label{etz1}
\end{equation}
\ \\
 \ \\
 In the Poisson REM of section \ref{sub:Definition-of-Poisson-REM},
with intensity (\ref{eq:energy_dist}) 
\[
\rho(E)=2^{N}P(E)
\]
that we consider in the present paper the same generating function
is given by 
\begin{equation}
\left\langle e^{-tZ}\right\rangle _{\text{Poisson}}=\exp\left[\int(e^{-te^{-\beta E}}-1)\ \rho(E)\ dE\right]\label{etz2}
\end{equation}
In the low temperature phase, to leading order, one can replace $\rho(E)$
by an exponential approximation (\ref{eq:rho_in_epsilon_form})

\[
\rho(E)=2^{N}P(E)\simeq Ae^{\alpha(E-E_{0})}
\]
and (\ref{etz2}) gives 
\begin{equation}
\left\langle e^{-tZ}\right\rangle _{\text{Poisson}}=\exp\left[\frac{Ae^{-\alpha E_{0}}}{\beta}\Gamma\left(-\frac{\alpha}{\beta}\ \right)\ t^{\frac{\alpha}{\beta}}\right]=\exp\left[-C\; t^{\frac{\alpha}{\beta}}\right]\label{etz3}
\end{equation}
where 
\[
C=-\frac{A}{\beta}\Gamma\left(-\frac{\alpha}{\beta}\ \right)e^{-\alpha E_{0}}\ .
\]
(Note that $C>0$ as the approximation (\ref{eq:rho_in_epsilon_form})
is only valid in the low temperature phase i.e. when $\alpha<\beta$
that is when only energies close to the ground state matter). Using
the fact that $P(E)$ is normalized one has 
\[
\log\left[\int e^{-te^{-\beta E}}\ P(E)\ dE\right]=\int(e^{-te^{-\beta E}}-1)\ P(E)\ dE-\frac{1}{2}\left[\int(e^{-te^{-\beta E}}-1)\ P(E)\ dE\right]^{2}+\cdots
\]
and one can see that 
\[
\left\langle e^{-tZ}\right\rangle _{\text{REM}}-\left\langle e^{-tZ}\right\rangle _{\text{Poisson}}\simeq-\frac{C^{2}}{2^{N+1}}t^{\frac{2\alpha}{\beta}}\exp\left[-C\; t^{\frac{\alpha}{\beta}}\right]\ .
\]
Then using the formula (\ref{Znu}) one get for $0<m<\mu=\alpha/\beta$
\[
\langle Z^{m}\rangle_{\text{Poisson}}\simeq C^{\frac{m\beta}{\alpha}}\;\frac{\beta}{\alpha}\frac{\Gamma(-\frac{\beta m}{\alpha})}{\Gamma(-m)}\ \ \ \ \ \ ;\ \ \ \ \ \ \langle Z^{m}\rangle_{\text{REM}}-\langle Z^{m}\rangle_{\text{Poisson}}\simeq-\frac{1}{2^{N+1}}C^{\frac{m\beta}{\alpha}}\;\frac{\beta}{\alpha}\frac{\Gamma(2-\frac{\beta m}{\alpha})}{\Gamma(-m)}\ .
\]
We see that the difference has an extra factor $2^{-N}$ which makes
the original REM and the Poisson version { coincide} to all orders
in a $1/N$ expansion.

\section{$1/N$ corrections at a saddle point}

In this appendix we derive a general formula for the $1/N$ corrections
of an arbitrary observable $H(x)$ at a saddle point. Suppose that
we want to evaluate, for large $N$, a ratio of the form 
\begin{equation}
\langle H(x)\rangle=\frac{\int{\rm d}x\; e^{NF(x)}\; G(x)\; H(x)}{\int{\rm d}x\; e^{NF(x)}\; G(x)}
\end{equation}
by a saddle point method. One has first to locate the saddle point
$x_{c}$ which satisfies 
\[
F'(x_{c})=0\ .
\]
Then by expanding $F,G,H$ around $x_{c}$ one finds

\begin{equation}
\langle H(x)\rangle=H(x_{c})+\frac{1}{N}\left[\left(\frac{F'''(x_{c})}{2F''(x_{c})^{2}}-\frac{G'(x_{c})}{G(x_{c})\; F''(x_{c})}\right)H'(x_{c})-\frac{1}{2F''(x_{c})}H''(x_{c})\right]+O\left(\frac{1}{N^{2}}\right)\ .
\end{equation}

One can rewrite this expression as 
\begin{equation}
\langle H(x)\rangle=\langle H(x_{c}+\eta)\rangle_{\eta}+O\left(\frac{1}{N^{2}}\right)
\end{equation}
where 
\begin{equation}
\langle\eta\rangle_{\eta}=\frac{1}{N}\left[\frac{F'''(x_{c})}{2F''(x_{c})^{2}}-\frac{G'(x_{c})}{G(x_{c})\; F''(x_{c})}\right]\,,\label{AP1}
\end{equation}
\begin{equation}
\langle\eta^{2}\rangle_{\eta}=-\frac{1}{N}\left[\frac{1}{F''(x_{c})}\right]\,.\label{AP2}
\end{equation}

Formulas would remain unchanged if the integration was not along the
real axis but along any path in the complex plane. In section \ref{saddlep}
we in fact use them along contours in the complex plane. \ \\
 \ \\

\bibliographystyle{elsarticle-num}

\bibliography{REM-finite-size-Dec14}

\end{document}